\pdfoutput=1
\documentclass[11pt,twoside,a4paper,cmspaper,final,collab]{cms-tdr}
\def\svnVersion{6002674}\def\svnDate{2025/01/01}\def\cmsCernNoTag{CERN-EP-2024-129}\def\cmsCernDate{\today}\def\cmsMessage{Published in Physics Letters B as \href{http://dx.doi.org/10.1016/j.physletb.2024.139231}{\doi{10.1016/j.physletb.2024.139231}.}}
\begin{document}\cmsNoteHeader{SMP-24-001}

\newcommand{\pb}{\unit{pb}}
\newcommand{\WWpm}{\ensuremath{\PWp\PWm}\xspace}
\newcommand{\WW}{\ensuremath{\PW\PW}\xspace}
\newcommand{\WZ}{\ensuremath{\PW\PZ}\xspace}
\newcommand{\ZZ}{\ensuremath{\PZ\PZ}\xspace}
\newcommand{\mll}{\ensuremath{m_{\ell\ell}}\xspace}
\newcommand{\mlll}{\ensuremath{m_{\ell\ell\ell}}\xspace}
\newcommand{\ptlmax}{\ensuremath{\pt^{\ell\text{\,max}}}\xspace}
\newcommand{\ptlmin}{\ensuremath{\pt^{\ell\text{\,min}}}\xspace}
\newcommand{\ptll}{\ensuremath{\pt^{\ell\ell}}\xspace}
\newcommand{\jet}{\ensuremath{\mathrm{j}}}
\newcommand{\DeepJet}{{\textsc{DeepJet}}\xspace}
\newcommand{\BFWlept}{{\mathcal{B}}(\PW\to\ell\nu)}
\newcommand{\NJ}{\ensuremath{N_{\mathrm{j}}}\xspace}
\newcommand{\tVx}{\ensuremath{\PQt\PV\mathrm{x}}\xspace}
\newcommand{\MATRIX}{\textsc{matrix}}
\newcommand{\pminnlo}{\POWHEG~\textsc{MiNNLO}\xspace}
\newlength\cmsTabSkip\setlength{\cmsTabSkip}{1ex}
\ifthenelse{\boolean{cms@external}}{
  \providecommand{\maybeCmsTable}[1]{\resizebox{0.49\textwidth}{!}{#1}}
  }{
  \providecommand{\maybeCmsTable}[1]{{#1}}
}

\cmsNoteHeader{SMP-24-001}

\title{Measurement of inclusive and differential cross sections for \texorpdfstring{$\PWp\PWm$}{W+W-} production in proton-proton collisions at \texorpdfstring{$\sqrt{s} = 13.6\TeV$}{sqrt(s) = 13.6 TeV}}

\date{\today}

\abstract{
Measurements at $\sqrt{s}=13.6\TeV$ of the opposite-sign \PW boson pair production cross section in proton-proton collisions are presented. The data used in this study were collected with the CMS detector at the CERN LHC in 2022, and correspond to an integrated luminosity of 34.8\fbinv. Events are selected by requiring one electron and one muon of opposite charge. A maximum likelihood fit is performed on signal- and background-enriched data categories defined by the flavour and charge of the leptons, the number of jets, and number of jets originating from \PQb quarks. The overall sensitivity is significantly better than that of previous results with a similar integrated luminosity. The improvement comes from a more refined control of experimental uncertainties and an improved fit strategy. An inclusive \WWpm production cross section of  $125.7 \pm 5.6$ pb is measured, in agreement with standard model predictions. Cross sections are also reported in a fiducial region close to that of the detector acceptance, both inclusively and differentially, as a function of the jet multiplicity in the event. For the first time in proton-proton collisions, \WW events with zero, one, and at least two jets are studied simultaneously and compared with recent theoretical predictions.}

\hypersetup{
pdfauthor={CMS Collaboration},
pdftitle={Measurement of inclusive and differential cross sections for W+W- production in proton-proton collisions at sqrt(s)=13.6 TeV},
pdfsubject={CMS},
pdfkeywords={CMS,  W boson pairs}}

\maketitle 

\section{Introduction}

The measurement of \PW boson pair (\WWpm or simply \WW) production cross sections is an important test of the standard model (SM). The  \WW production at hadron colliders is sensitive to the properties of electroweak (EW) boson self-interactions and provides a test of the predictions of perturbative quantum chromodynamics (QCD) and the EW theory. It also constitutes a large background in the measurement of Higgs boson production, as well as in searches for physics beyond the SM. It is important to model the production of \WW{+}jets accurately in event generators. This process is also background in $\ttbar$ production studies and in searches for physics beyond the SM with charged leptons and jets in the final state~\cite{CMS:2018kag}. The \WW cross section has been measured inclusively and differentially in proton-antiproton ($\Pp\PAp$) collisions at $\sqrt{s} = 1.96\TeV$~\cite{Abazov:2009ys,D0:2013bwr,Aaltonen:2009aa} and in proton-proton ($\Pp\Pp$) collisions at 7\TeV~\cite{ATLAS:2012mec,Chatrchyan:2013yaa}, 8\TeV~\cite{Khachatryan:2015sga, Aad:2016wpd,ATLAS:2016agv}, and 13\TeV~\cite{ATLAS:2019nkz,CMS:2020mxy,ATLAS:2021jgw}, and agrees with the SM predictions within uncertainties.

This letter presents the first measurements at $\sqrt{s} = 13.6\TeV$ of the inclusive and differential $\WWpm\to\Pe^\pm\nu\PGm^\mp\nu$ production cross sections. Data from pp collisions recorded by the CMS experiment in 2022 are used, corresponding to an integrated luminosity of $34.8\fbinv$. The SM production of \WW pairs proceeds mainly through three processes: the dominant $\cPq\cPaq$ annihilation process; the $\Pg\Pg\to\WW$ process; and the Higgs boson process $\PH\to\WW$, which is approximately ten times smaller than the sum of the other two processes and is considered as a background in this analysis. A calculation of the \WW production cross section in $\Pp\Pp$ collisions at $\sqrt{s} = 13.6\TeV$ using \textsc{MATRIX}~v2.1.0~\cite{Grazzini:2017mhc,Grazzini:2019jkl} yields the value $125.8\pm3.9\pb$ at next-to-next-to-leading order (NNLO) in QCD and next-to-leading order (NLO) in EW. The calculation uses the parton distribution function (PDF) NNPDF 3.0~\cite{Ball:2017nwa}. The uncertainties reflect the dependence of the calculation expansion on the QCD factorization and renormalization scales, and the PDFs. The $\Pg\Pg\to\WW$ process contributes about 3\%, after correcting by a factor of 1.4, which comes from the ratio of the $\Pg\Pg\to\WW$ cross section at NLO to the same cross section at leading order (LO)~\cite{Caola:2015rqy}. The most important background processes are (i) top quark production (\ttbar and $\cPqt\PW$), (ii) Drell--Yan production of leptonically decaying tau lepton pairs, and (iii) nonprompt leptons originating from leptonic decays of heavy quarks, hadrons misidentified as leptons, and electrons from photon conversions.

In this analysis, events are selected with one electron and one muon of opposite charge, including those originating from leptonic tau decays. Events with two charged leptons of the same flavour are not considered in the analysis because of the much higher background from Drell--Yan events. To measure the  \WW signal and estimate the most important background processes, events are further categorized into a signal region (SR) and control regions (CRs) defined by kinematic requirements on the flavour and charge of the leptons, the number of jets, and the number of jets originating from \PQb quarks (b jets). The inclusive  \WW cross section is extracted from a profile maximum likelihood fit~\cite{CMS:2024onh} to the event yields in the SR and CRs. The overall sensitivity is about 25\% better than the results reported in Ref.~\cite{CMS:2020mxy} at $\sqrt{s}=13\TeV$ with a similar integrated luminosity. The improvement comes from several reduced experimental uncertainties and an improved fit strategy. The  \WW cross section is also measured in a fiducial region close to the detector acceptance, both inclusively and differentially, as a function of the jet multiplicity in the event. Events consistent with \WW production and up to two reconstructed jets inclusively are studied in $\Pp\Pp$ collisions and compared with recent theoretical predictions.

\section{The CMS detector and simulated samples}

The central feature of the CMS apparatus is a superconducting solenoid of 6\unit{m} internal diameter, providing a magnetic field of 3.8\unit{T}. A silicon pixel and strip tracker, a lead tungstate crystal electromagnetic calorimeter (ECAL) with a silicon strip preshower detector placed at the front faces of the two endcap calorimeters, and a brass and scintillator hadron calorimeter (HCAL), each composed of a barrel and two endcap sections, are installed inside the solenoid. Forward calorimeters extend the pseudorapidity ($\eta$) coverage provided by the barrel and endcap detectors. Muons are detected in gas-ionization chambers embedded in the steel flux-return yoke outside the solenoid. A more detailed description of the CMS detector, together with a definition of the coordinate system and the relevant kinematic variables, is presented in Refs.~\cite{CMS:2008xjf,CMS:2023gfb}.

Events of interest are selected using a two-tiered trigger system. The first level, composed of custom hardware processors, uses information from the calorimeters and muon detectors to select events at a rate of around 100\unit{kHz} within a fixed latency of 4\mus~\cite{CMS:2020cmk}. The second level, known as the high-level trigger, consists of a farm of processors running a version of the full event reconstruction software optimized for fast processing, and reduces the event rate to around 5\unit{kHz} before data storage~\cite{CMS:2016ngn}.

Several Monte Carlo (MC) event generators are used to simulate the signal and background processes. The simulated samples are used to optimize the event selection, evaluate selection efficiencies and systematic uncertainties, and compute expected yields. The quark-initiated nonresonant \WW, other heavy vector dibosons, \ttbar and $\cPqt\PW$ processes are simulated with \POWHEG v2~\cite{Nason:2004rx,Frixione:2007vw,powheg:2010,Nason:2013ydw,Alioli:2008gx,Alioli:2008tz} at NLO accuracy for inclusive production. The \MADGRAPH 2.9.9 generator~\cite{Alwall:2014hca,Alwall:2007fs,Frederix:2012ps,Artoisenet:2012st} is used to simulate gluon-initiated \WW production at LO accuracy. The production of a Higgs boson is also generated with \POWHEG, where the Higgs bosons are decayed with \textsc{JHUGen}~v5.2.5~\cite{jhu2014,jhu2020}. Production of \PZ{+}jets, \PW{+}jets, $\PZ\PGg$, $\PW\PGg$, $\ttbar\PW$, $\ttbar\PZ$, $\ttbar\WW$, and triple vector boson ($\PV\PV\PV$, $\PV=\PW/\PZ$) events is simulated at NLO accuracy in QCD using the \MGvATNLO generator.

The NNPDF 3.1 NNLO~\cite{Ball:2017nwa} PDFs are used for simulating all samples. For all processes, the underlying event, parton showering (PS), and hadronization are simulated using $\PYTHIA$ 8.306~\cite{Sjostrand:2015}. The modelling of the underlying event is performed using the CP5~\cite{Sirunyan:2019dfx} tune. All MC generated events are processed through a simulation of the CMS detector based on \GEANTfour~\cite{Agostinelli:2002hh} and are reconstructed with the same algorithms as used for the data. Additional $\Pp\Pp$ interactions in the same and nearby bunch crossings, referred to as pileup, are also simulated. The distribution of the number of pileup interactions in the simulation is adjusted to match the one observed in the data, with an average pileup of about 50 interactions per bunch crossing.

\section{Object reconstruction}

A particle-flow (PF) algorithm~\cite{CMS:2017yfk} reconstructs and identifies each individual particle in an event, with an optimized combination of information from the various elements of the CMS detector. The energy of photons is obtained from the ECAL measurement. The energy of electrons is determined from a combination of the electron momentum at the primary interaction vertex as determined by the tracker, the energy of the corresponding ECAL cluster, and the energy sum of all bremsstrahlung photons spatially compatible with originating from the electron track. The energy of muons is obtained from the curvature of the corresponding track. The energy of charged hadrons is determined from a combination of their momentum measured in the tracker and the matching ECAL and HCAL energy deposits, corrected for the response function of the calorimeters to hadronic showers. Finally, the energy of neutral hadrons is obtained from the corresponding corrected ECAL and HCAL energy deposits. The primary vertex (PV) is taken to be the vertex corresponding to the hardest scattering in the event, evaluated using tracking information alone, as described in Section 9.4.1 of Ref.~\cite{CMS-TDR-15-02}.

For each event, hadronic jets are clustered from these reconstructed particles using the anti-\kt algorithm~\cite{Cacciari:2008gp,Cacciari:2011ma} with a distance parameter of 0.4. The jet momentum is determined as the vectorial sum of all particle momenta in the jet, and is found from simulation to differ, on average, less than 5-10\% from the true momentum over the whole transverse momentum (\pt) spectrum and detector acceptance. Jet energy corrections are derived from simulation to bring, on average, the measured response of jets close to that of generator-level jets. In situ measurements of the momentum balance in dijet, $\gamma${+}jet, \PZ{+}jet, and multijet events are used to account for any residual differences in the jet energy scale between data and simulation~\cite{CMS:2016lmd}. The jet energy resolution amounts typically to 15--20\% at 30\GeV, 10\% at 100\GeV, and 5\% at 1\TeV~\cite{CMS:2016lmd}. Selection criteria are applied to remove jets potentially dominated by anomalous contributions from various subdetector components or reconstruction failures~\cite{Sirunyan:2019kia,CMS-PAS-JME-16-003}. Pileup interactions can produce additional tracks and calorimetric energy depositions to the jet momentum. The pileup-per-particle identification algorithm (PUPPI)~\cite{Sirunyan:2020foa,Bertolini:2014bba} is used to mitigate the effect of pileup at the reconstructed particle level, making use of local shape information, event pileup properties, and tracking information. A local shape variable is defined, which distinguishes between collinear and soft diffuse distributions~\cite{Sirunyan:2020foa} of other particles surrounding the particle under consideration. The former is attributed to particles originating from the hard scattering and the latter to particles originating from pileup interactions. Charged particles identified to be originating from pileup vertices are discarded. For each neutral particle, a local shape variable is computed using the surrounding charged particles compatible with the primary vertex within the tracker acceptance ($\abs{\eta} < 2.5$), and using both charged and neutral particles in the region outside of the tracker coverage. The momenta of the neutral particles are then rescaled according to their probability to originate from the primary interaction vertex deduced from the local shape variable, superseding the need for jet-based pileup corrections~\cite{Sirunyan:2020foa}. The jet multiplicity \NJ counts the number of reconstructed jets with $\pt>30\GeV$ and $\abs{\eta} < 2.5$. Jets falling within $\Delta R = \sqrt{\smash[b]{(\Delta\eta)^2+(\Delta\phi)^2}} < 0.4$ (where $\phi$ is azimuthal angle in radians) of a selected charged lepton, as described below, are discarded.
 
The missing transverse momentum vector, \ptvecmiss, is computed as the negative vector $\pt$ sum of all the PF candidates in an event, and its magnitude is denoted as \ptmiss~\cite{Sirunyan:2019kia}. The \ptvecmiss is corrected to account for corrections to the energy scale of the reconstructed jets in the event. The PUPPI algorithm is applied to reduce the pileup dependence of the \ptvecmiss observable. The \ptvecmiss is computed from the PF candidates weighted by their probability to originate from the PV~\cite{Sirunyan:2019kia}. 

Jets with $\pt>20\GeV$ and $\abs{\eta}<2.5$ originating from \PQb quarks are identified (``tagged'') by a multivariate algorithm called \DeepJet~\cite{Bols:2020bkb}. For the chosen working point, the efficiency to select \PQb\ jets is about 90\% and the rate for incorrectly tagging jets originating from the hadronization of gluons or $\cPqu$, $\cPqd$, $\cPqs$ quarks is about 10\%. The rate for mistagging jets originating from the hadronization of $\cPqc$ quarks is about 50\%.

Electrons and muons are reconstructed by associating a track reconstructed in the tracking detectors with either a cluster of energy in the ECAL~\cite{CMS:2020uim} or a track in the muon system~\cite{Sirunyan:2018fpa}. Electron and muon candidates must pass certain identification criteria to be further selected in the analysis. A ``loose'' identification is defined by requiring $\pt > 10\GeV$ and $\abs{\eta} < 2.5$ (2.4) for electrons (muons). A ``tight'' selection is required as a last step of the lepton selection following the definitions provided in Refs.~\cite{CMS:2020uim,Sirunyan:2018fpa}, including requirements on the impact parameter of the candidates with respect to the PV and their isolation with respect to other particles in the event~\cite{Sirunyan:2018egh}. 

\section{Event selection}\label{sec:selection}

Collision events were collected using a single-electron (single-muon) trigger that requires the presence of an isolated lepton with $\pt>27$ (24)\GeV. In addition, a set of dilepton triggers with a threshold of $23\GeV$ for the leading lepton and with a threshold of 8\GeV for the subleading lepton are used. This ensures a trigger efficiency above 99\% for events that satisfy the offline selection.

In the analysis, the SR requirements are applied to isolate the signal topology while reducing the background processes. The CR requirements are applied to select regions dominated by background processes to constrain them. The main analysis strategy is to simultaneously fit all these regions to measure the signal processes and to control the normalization of the background components. 

The leptonic final state in the $\WW\to\Pe^\pm\nu\PGm^\mp\nu$ channel consists of two oppositely charged, high-$\pt$, isolated leptons. Selected events contain exactly one electron and one muon with opposite charge, including those originating from leptonic tau decays. The leading ($\ptlmax$) and subleading ($\ptlmin$) lepton $\pt$ values are required to be larger than 25 and 20\GeV, respectively. Events containing any additional loosely identified lepton with $\pt>10\GeV$ are rejected to suppress backgrounds from $\WZ$ and $\ZZ$ processes. The dilepton invariant mass (\mll) is required to be larger than 85\GeV to reject $\PZ\to\tau^+\tau^-$ and Higgs boson background events. To reduce top quark background contributions, events with one or more \PQb-tagged jets are rejected.

Several orthogonal background-enriched CRs are used to select event samples enriched with top quark (referred to as ``one $\PQb$ tag'' and ``two $\PQb$ tags''), $\PZ\to\tau^+\tau^-$ (referred to as ``$\PZ\to\tau\tau$''), and nonprompt-lepton background  (referred to as ``same-sign'') events. The top quark CRs are defined by requiring the  \WW SR selection with the exception of the \PQb\ jet veto requirement, which require either exactly one or two \PQb\ jets. The $\PZ\to\tau\tau$ CR is defined by inverting the $\mll$ requirement, and by requiring $\ptll<30\GeV$. Finally, the nonprompt-lepton CR is defined by requiring a total lepton charge of $\pm2$. Table~\ref{tab:selectioncuts2l} summarizes the event selection used in these regions. All regions are split in events with 0, 1, 2, and $\geq3$ reconstructed jets.

\begin{table*}[htbp]
\begin{center}
\topcaption{Summary of the requirements defining the  \WW SR and dilepton CRs.\label{tab:selectioncuts2l}}
  \begin{tabular}{ l cc cc }
  \hline
 Quantity & $\PW\PW$ & One/two \PQb tags & $\PZ\to\tau\tau$ & Same-sign \\
  \hline
Number of tight leptons      & \multicolumn{4}{c}{Strictly 2}  \\
Additional loose leptons     & \multicolumn{4}{c}{0} \\
Lepton charges               & \multicolumn{3}{c}{Opposite}    & Same  \\
$\ptlmax$                    & \multicolumn{4}{c}{$>$25\GeV} \\
$\ptlmin$                    & \multicolumn{4}{c}{$>$20\GeV} \\
$\mll$                       & $>$85\GeV & $>$85\GeV & $<$85\GeV & $>$85\GeV \\
$\ptll$                      & \NA & \NA & $<$30\GeV & \NA \\
Number of \Pb-tagged jets    & 0 & $1/2$ & 0 & 0 \\
$\NJ$                        & \multicolumn{4}{c}{$0/1/2/\geq3$} \\
  \hline
  \end{tabular}
\end{center}
\end{table*}

In addition, two CRs enriched in $\WZ\to3\ell\nu$ and $\ZZ\to4\ell$ events are selected in the data, as described in Table~\ref{tab:selectioncuts3l}. The $\WZ\to3\ell\nu$ region is selected by requiring three leptons with a total charge $\pm1$. One of the oppositely charged same-flavour leptons from the $\PZ$ boson candidate is required to have $\pt>25\GeV$ and the other $\pt>10\GeV$, and the invariant mass of the pair must satisfy $\abs{\mll - m_{\PZ}}<15\GeV$. In candidate events with three same-flavour leptons, the oppositely charged leptons with the invariant mass closest to the nominal $\PZ$ boson mass $m_{\PZ}$~\cite{Workman:2022ynf} is selected as the $\PZ$ boson candidate. The third lepton with $\pt>20\GeV$ is associated with the $\PW$ boson. The trilepton invariant mass ($\mlll$) is required to exceed 100\GeV. The $\ptmiss$ associated with the undetected neutrinos is required to be greater than 30\GeV. Finally, to reduce top quark background contributions, events with one or more jets tagged as \PQb~jets are rejected. The $\ZZ\to4\ell$ region is selected by requiring four leptons, where at least one lepton must have $\pt>25\GeV$. Lepton pairs originating from hadronic decays are removed by requiring that all opposite-sign lepton pairs in the candidate have $\mll>5\GeV$. The $\PZ$ boson candidates are built from two oppositely-charged leptons of the same flavour. Events with two of such pairs are retained. Finally, to emulate the  \WW SR, events with one or more \PQb~tagged jets are rejected.

\begin{table*}[htbp]
  \begin{center}
  \caption{Summary of the requirements defining the $\WZ$ and $\ZZ$ CRs.\label{tab:selectioncuts3l}}
  \begin{tabular} {lcc}
  \hline
  Variable & $\WZ$ & $\ZZ$ \\
  \hline
Number of tight leptons          & Strictly 3 & Strictly 4 \\
Additional loose leptons         & \multicolumn{2}{c}{0} \\
Lepton $\pt$                     & ${>}25/10/20\GeV$ & ${>}25/20/10/10\GeV$ ($\pt$ ordered) \\
$\abs{\mll - m_{\PZ}}$           & ${<}15\GeV$ & ${<}15\GeV$  (both pairs) \\
$m_{3\ell}$                      & ${>}100\GeV$    &  \NA  \\
$m_{4\ell}$                      &  \NA  & ${>}150\GeV$ \\
$\ptmiss$                        & ${>}30\GeV$    &  \NA  \\
Number of \Pb-tagged jets        & \multicolumn{2}{c}{0} \\
  \hline
  \end{tabular}
  \end{center}
\end{table*}

\section{Background estimation}

A combination of methods based on CRs in data and simulation is used to estimate background contributions. The main background processes are normalized by comparing the predicted yields to data in several CRs, defined to be as close as possible to the SR.

The largest background contribution comes from top quark production, predominantly from \ttbar and $\PQt\PW$ events. This contribution arises when \PQb~jets are not tagged either because they fall outside the kinematic region where tagging is possible or because they receive low scores from the \DeepJet \PQb tagging algorithm. An estimate of the top quark background is obtained in each individual reconstructed jet multiplicity bin using the CRs explained in Section~\ref{sec:selection}. Following the procedure used in a measurement of the $\ttbar$ production cross section~\cite{CMS:2023qyl}, two CRs requiring exactly one and exactly two $\PQb$-tagged jets are defined. By considering events in these two regions separately, the impact of the uncertainties in $\PQb$  tagging efficiency is reduced by about 50\%. The next largest background contribution comes from the Drell--Yan process from $\PZ\to\PGt\PGt$ production with both $\PGt$ leptons decaying leptonically. The CR described in Section~\ref{sec:selection} is used to estimate the normalization of this process.

The next most important background contribution comes from events with nonprompt leptons. The simulation alone cannot be used for an accurate estimate of this contribution, but it can be used in conjunction with the data yields in the CR  to determine a background estimate. A ``pass-fail'' control sample is defined by one lepton that passes the lepton selection criteria and another that fails the criteria but passes  looser criteria. The misidentification rate $f$ for a jet that satisfies loose lepton requirements and also passes the standard lepton requirements is determined using a sample dominated by multijet events with nonprompt leptons. This misidentification rate is parameterized as a function of lepton \pt and $\eta$ and used to compute weights $f/(1-f)$ in the pass-fail sample that are used to determine the contribution of nonprompt leptons in the signal region~\cite{Sirunyan:2018egh,HiggsWW2016}. The systematic uncertainty to estimate this background is dominated by the uncertainty in the determination of $f$, which has two sources: the dependence of $f$ on the sample composition, and the method. The first source is estimated by modifying the jet $\pt$ threshold in the QCD multijet sample and by requiring one \PQb~jet, which changes the composition. The uncertainty in the method is obtained from a closure test, where $f$ is derived from simulated QCD multijet events and applied to simulated samples to predict the number of background events. The limited sample size of the control sample is also taken into account. In addition, the same-sign CR, enhanced in nonprompt lepton background events, is used to estimate in situ the normalization of this background, as described in Section~\ref{sec:selection}.

The normalizations of the $\WZ\to3\ell\nu$ and $\ZZ\to4\ell$ background processes are estimated using the dedicated CRs described in the previous section. The remaining minor sources of background, including triboson final states, top quarks in association with vector bosons, and Higgs-mediated \WW production, are evaluated using simulations normalized to the most precise theoretical cross sections available.

\section{Systematic uncertainties}

Multiple sources of systematic uncertainty are considered in the measurement of the inclusive and differential cross sections. The uncertainties are included through nuisance parameters in the profile likelihood fit~\cite{CMS:2024onh,Cowan:2010js}, described in Section~\ref{sec:results}. They can affect either the normalization or both the normalization and the shape of the distributions of the signal or background processes. For each source, the predicted event samples are used to construct corresponding template histograms that describe the expected signal and background distributions for a specific nuisance parameter variation. In the fit of the templates to the data, the best values for both the measured cross section and all nuisance parameters are determined simultaneously.

The uncertainty in the integrated luminosity measurements for the data used in this analysis is 1.4\%~\cite{CMS-PAS-LUM-22-001}. The uncertainty in the simulation of pileup events is evaluated by varying the total inelastic pp cross section of 69.2\unit{mb} by $\pm$5\%, which has an impact in the expected signal and background yields of about 1\%.

Discrepancies in the lepton reconstruction and identification efficiencies between data and simulation are corrected by applying scale factors to all simulation samples. These scale factors, which depend on the $\pt$ and $\eta$ for both electrons and muons, are determined using $\PZ \to \Pell\Pell$ events in the $\PZ$ boson peak region that were recorded with independent triggers~\cite{Sirunyan:2018fpa,CMS:2020uim,Sirunyan:2019bzr}, with associated uncertainties of about 0.5\% for both electrons and muons. The uncertainty in the determination of the trigger efficiency leads to an uncertainty smaller than 1\% in the expected signal yield. The lepton momentum scale uncertainty is computed by varying the momenta of the leptons in simulation by their uncertainties, and repeating the analysis selection. The resulting uncertainties in the yields are about 0.1\% for both electrons and muons.

The uncertainty in the calibration of the jet energy scale and resolution affects the selection efficiency of the jet multiplicity requirement and the $\ptmiss$ measurement. These effects are estimated by changing the jet energy in the simulation up and down by one standard deviation. The uncertainty in the jet energy scale and resolution is 2--5\%, depending on the $\pt$ and $\eta$~\cite{CMS:2016lmd}, and the impact on the expected signal and background yields is 1--4\% depending on the reconstructed jet multiplicity bin. 

Scale factors are applied in the simulation for the \PQb tagging efficiency of individual \PQb jets and the mistagging efficiencies of \PQc quark, light-flavour, and gluon jets are measured using dedicated calibration samples~\cite{Sirunyan:2017ezt}.  The statistical uncertainties in the scale factors are uncorrelated between data-taking periods, while other uncertainties in the calibration method are treated as fully correlated. All uncertainties are treated as correlated among all regions in the fit. Each set of values results in uncertainties in the \PQb tagging efficiency of about 1--4\% to select events in the  \WW SR, and the impact on the expected signal and background yields is about 1\%.

The dominant theoretical uncertainties for simulated processes, both signal and background, correspond to missing higher-order QCD calculations, which are estimated by varying the QCD renormalization and factorization scales independently up and down by a factor of two from their nominal values. The largest cross section variation is taken as the uncertainty, excluding the two extreme variations where one scale is varied up and the other one down. The PDF and strong coupling $\alpS$ uncertainties are evaluated according to the procedure described in Ref.~\cite{PDFLHC}. The theoretical uncertainty due to the modelling of the PS is included for all the simulated samples. The uncertainty in the PS modelling is evaluated by varying the PS weights related to initial- and final-state radiation contributions computed by \PYTHIA on an event-by-event basis~\cite{Sjostrand:2015}, taking the uncertainty as the largest cross section variation on every analysis bin. The theoretical uncertainties are treated as fully correlated across bins for the distributions used to extract the results.

The branching fraction for leptonic decays of \PW bosons is $\BFWlept=0.1086\pm0.0009$~\cite{Workman:2022ynf}, and lepton universality is assumed to hold. This is used to quote the total inclusive cross section starting from the undecayed \PW bosons. The uncertainty coming from this branching fraction is not included in the uncertainty to estimate the inclusive cross section, since all theoretical predictions are computed using fully leptonic decays, and a simple numerical multiplication is applied to quote the total value. A summary of all other systematic uncertainties for inclusive signal strength ($\mu$), the inclusive cross section measurement divided by the predicted cross section, is given in Table~\ref{tab:wwSystematics}.

\begin{table}[htbp]
\centering
\topcaption{\label{tab:wwSystematics}
Systematic uncertainties in $\mu$, the measured inclusive cross section divided by the predicted cross section. Lepton experimental uncertainties encompass the effects of calibration of lepton momentum scale and resolution, as well as lepton trigger, reconstruction, identification, and isolation efficiencies. Jet experimental uncertainties encompass the effects of the jet energy scale and resolution. The limited sample size category is related to the finite number of MC events and data events used for data-driven background estimates.}
\begin{tabular}{lc}
  \hline
Uncertainty source &  $\Delta\mu$ \\
  \hline
Integrated luminosity    & 0.014 \\
Lepton experimental      & 0.019 \\
Jet experimental         & 0.008 \\
$\Pb$ tagging            & 0.012 \\
Nonprompt background     & 0.010 \\
Limited sample size      & 0.017 \\
Background normalization & 0.018 \\
Theory                   & 0.011 \\
Statistical              & 0.018 \\[\cmsTabSkip]

Total                    & 0.044 \\
  \hline
\end{tabular}
\end{table}

\section{Results}\label{sec:results}

The results include the  \WW total inclusive cross section, and both the inclusive and normalized differential cross sections measured as a function of the jet multiplicity in the event, in a fiducial region that is close to the detector acceptance. The cross sections are extracted from the data using a binned maximum likelihood fit~\cite{CMS:2024onh,Cowan:2010js} from the observed yields in $\NJ$ using the $\PW\PW$ SR and the one/two \PQb tags, $\PZ\to\PGt\PGt$, same-sign, $\WZ$ and $\ZZ$ CRs. In differential distributions, for each source of uncertainty, the impact in different bins is considered fully correlated. As described above, the normalization of the main background contributions are constrained for the extraction of results by the data in the CRs. In this fit, a signal strength parameter modifies the predicted signal yield defined by the central value of the theoretical cross section. The fitted value of the signal strength is expected to be close to unity if the SM is valid, and the measured cross section is the product of the signal strength and the theoretical cross section. All uncertainties considered in the analysis are treated as nuisance parameters in the fit. When building the likelihood function, each shape systematic uncertainty is associated to a nuisance parameter taken from a unit Gaussian distribution, which is used to interpolate or extrapolate using the specified histograms, and correlations between processes are also included. Statistical uncertainties in the signal and backgrounds are also treated as nuisance parameters, but are constrained by assigning a Poisson function to each analysis bin~\cite{Barlow:1993dm}. These constraints penalize the likelihood fit if the estimated nuisance parameters pull from their measured values.

The $\WWpm\to\Pe^\pm\PGn\PGm^\mp\PGn$ fiducial region is defined by a common set of kinematic requirements across the muon and electron final states, with all quantities evaluated at generator level after PS and hadronization, emulating the selection performed at the reconstruction level. The measured distributions, after subtracting the contributions from the background processes, are corrected for detector resolution effects and inefficiencies. The leptons at generator level are selected at the so-called ``dressed level'' by combining the four-momentum of each lepton after the final-state photon radiation with that of photons found within a cone of $\Delta R=0.1$ around the lepton~\cite{CMS-NOTE-2017-004,Sirunyan:2019bzr}. The fiducial region is defined by requiring two opposite-sign electrons and muons with $\abs{\eta}<2.5$, with a leading and trailing $\pt>25\GeV$ and $\pt>20\GeV$, respectively. Furthermore, $\mll$ is required to be larger than 85\GeV. The jets at the generator level are clustered from stable particles (\ie those with lifetimes greater than $0.3\times10^{-10}\unit{s}$), excluding neutrinos, using the anti-$\kt$ clustering algorithm with R = 0.4, and are required to have $\pt>30\GeV$ and $\abs{\eta}<2.5$. The jets within $\Delta R(\jet,\Pell)<0.4$ of the selected charged leptons are excluded. Electrons and muons produced in the decay of a $\PGt$ lepton contribute about 10\% of the signal sample and are excluded from the definition of the fiducial region, although they are treated as part of the signal component. All fiducial requirements are listed in Table~\ref{tab:fiducial}. Nonfiducial events, \ie events selected at the reconstructed level that do not satisfy the fiducial requirements, are included as background processes in the simultaneous fit.

\begin{table}[htbp]
  \centering
  \topcaption{
    Definition of the fiducial region.\label{tab:fiducial}
  }
  \maybeCmsTable{
  \begin{tabular}{lc}
    \hline
    Variable                               & Requirement \\
    \hline
    Lepton origin                          & Direct decay of a prompt $\PW$ boson \\
    Lepton definition                      & Dressed leptons ($\Pe^\pm\Pgm^\mp$) \\
    Leading lepton \pt                     & $\ptlmax > 25\GeV$ \\
    Trailing lepton \pt                    & $\ptlmin > 20\GeV$ \\
    Additional leptons                     & 0 \\
    $\abs{\eta}$ of leptons                & $\abs{\eta} < 2.5$ \\
    Dilepton mass                          & $\mll > 85\GeV$ \\
    Jet \pt                                & $\pt^\jet > 30\GeV$ \\
    $\abs{\eta}$ of jets                   & $\abs{\eta^\jet} < 2.5$ \\
    Jet-lepton removal                     & $\Delta R(\jet,\ell)>0.4$ \\
    \hline
  \end{tabular}
  }
\end{table}

To perform inclusive fiducial and normalized differential production cross section measurements, signal templates from different bins of the generator-level observable predicted by the event generator are built~\cite{CMS:2020mxy,CMS:2020gfh,CMS:2020hkd}. Each signal template is considered as a separate process in the simultaneous binned maximum likelihood fit; thus, the signal extraction and the unfolding of the detector effects are achieved in a single step. The ratios of the differential cross sections relative to the total fiducial cross section (normalized cross sections) and the inclusive fiducial cross section are evaluated simultaneously, and therefore, the systematic uncertainties in the individual cross sections are largely reduced. The total fiducial cross section and the fractions of events in two of the three jet multiplicity bins are freely floating in the fit. In this way, the normalized cross section has the same number of degrees of freedom as a differential measurement. The fraction of nonfiducial events and pileup jets for the differential measurements is below 1\%. The fraction of events in a different fiducial and reconstructed jet bin is relatively small, and is 5--15\% depending on the jet fiducial bin. Given that the bin migration effects are small, no regularization~\cite{Schmitt:2012kp,Cowan:2010js,Cowan} needs to be applied for these fits.

\begin{figure*}[htbp!]
\centering
\includegraphics[width=0.41\textwidth]{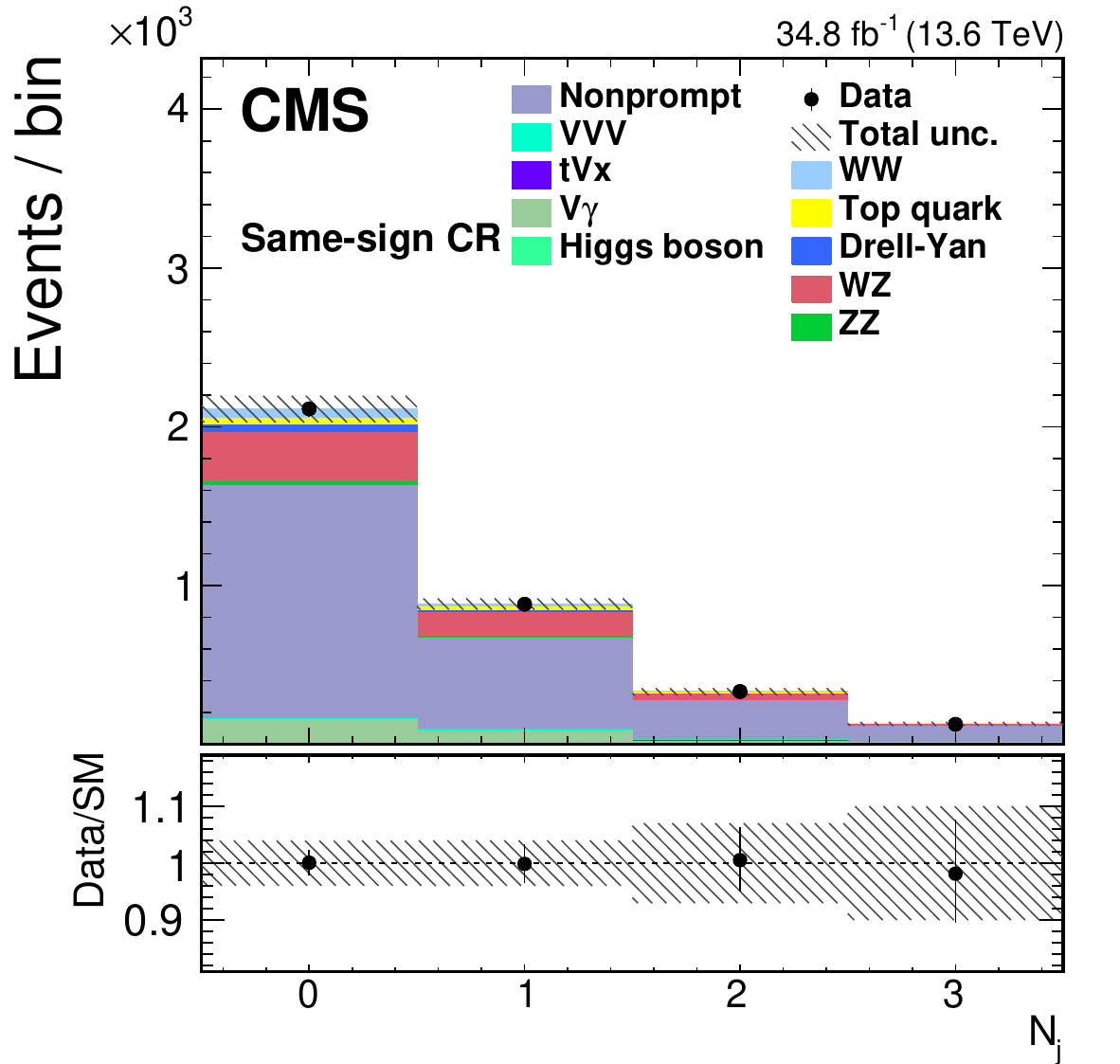}
\includegraphics[width=0.41\textwidth]{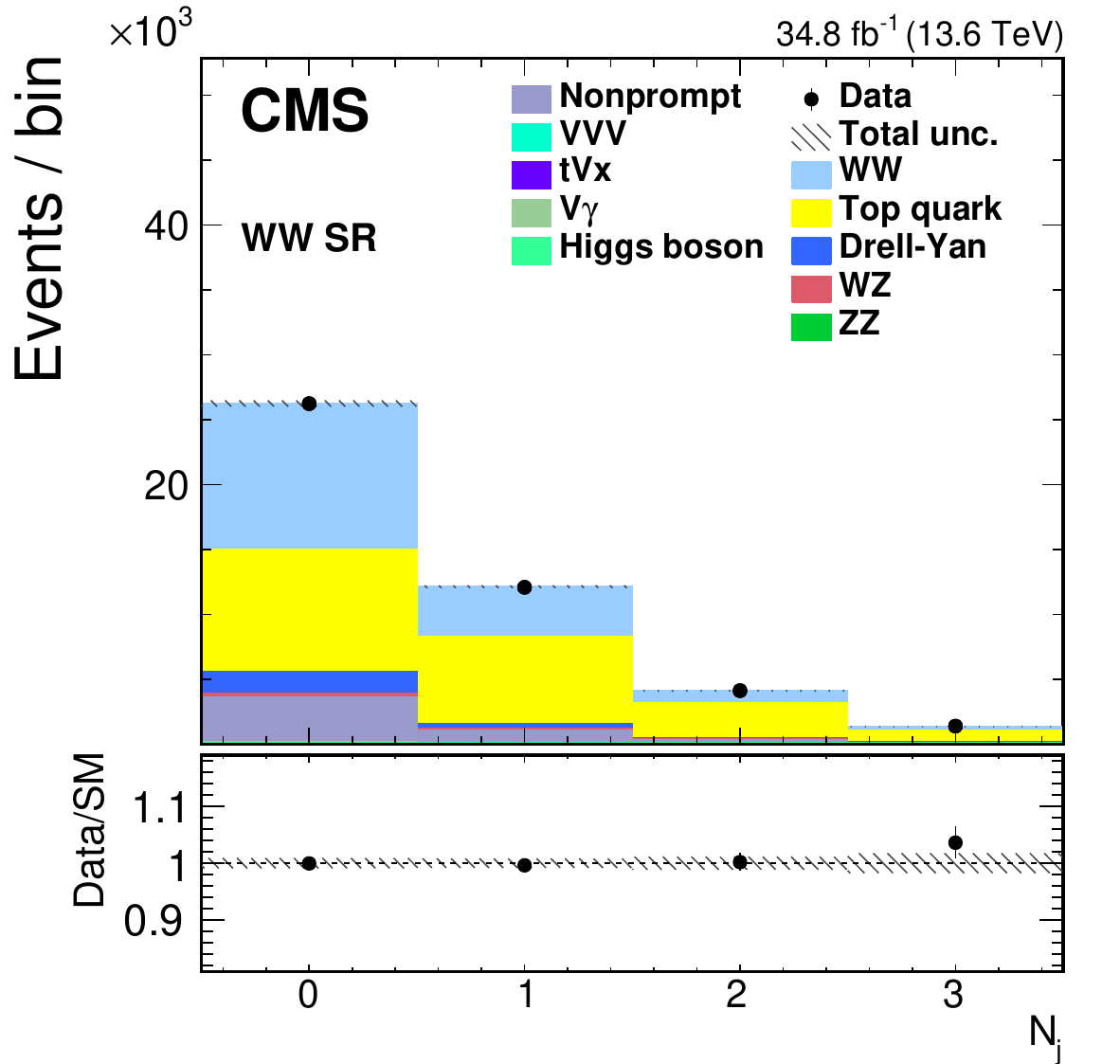}
\includegraphics[width=0.41\textwidth]{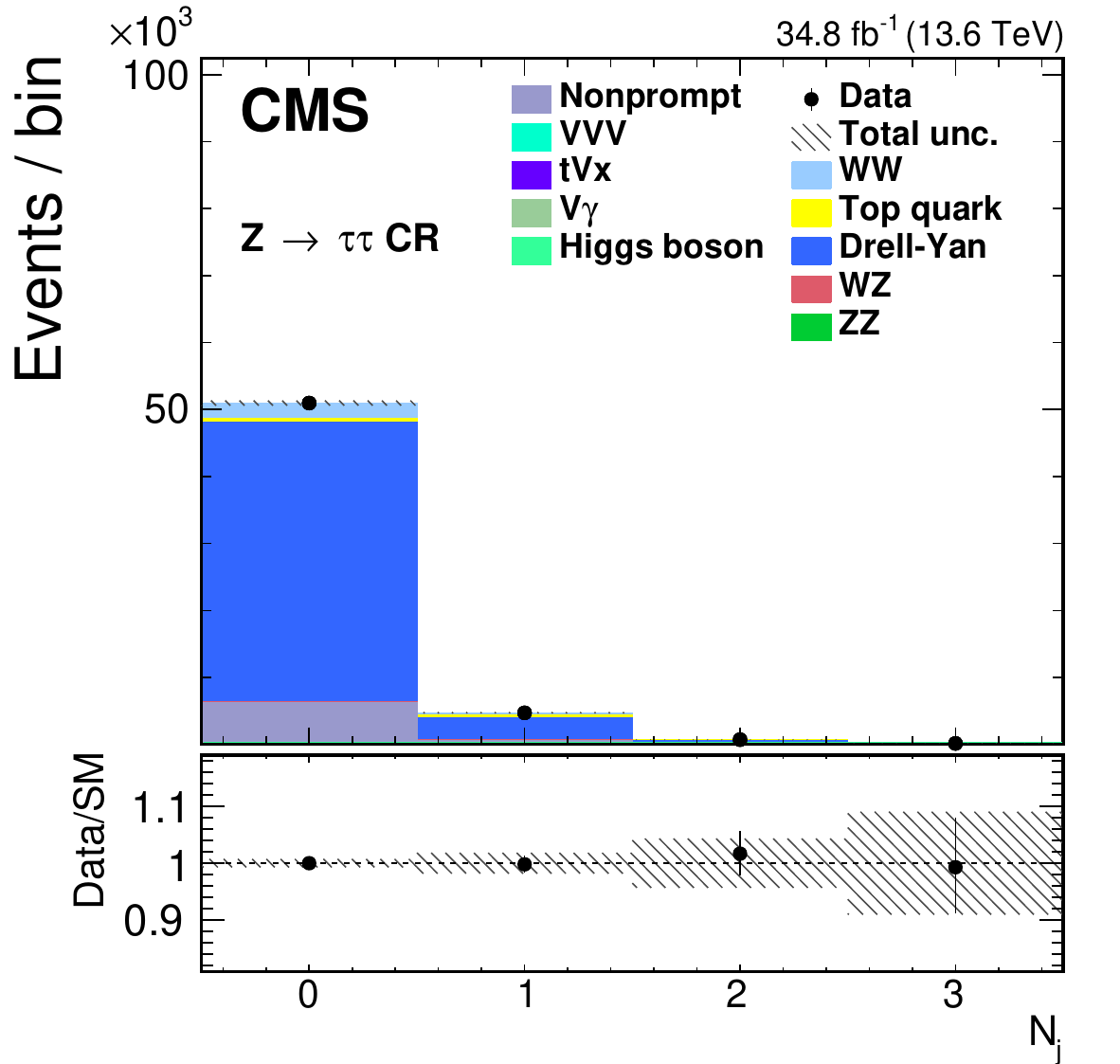}
\includegraphics[width=0.41\textwidth]{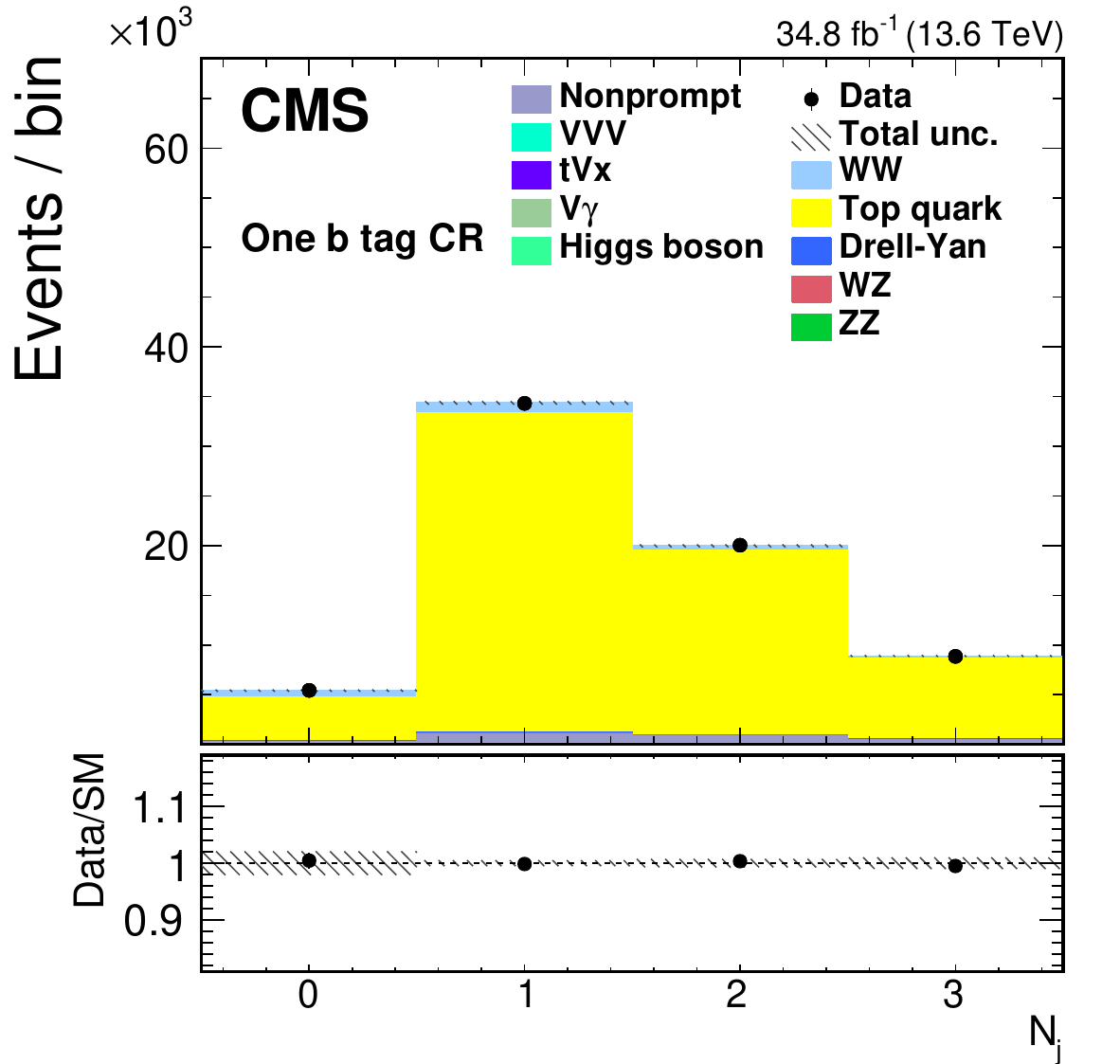}
\includegraphics[width=0.41\textwidth]{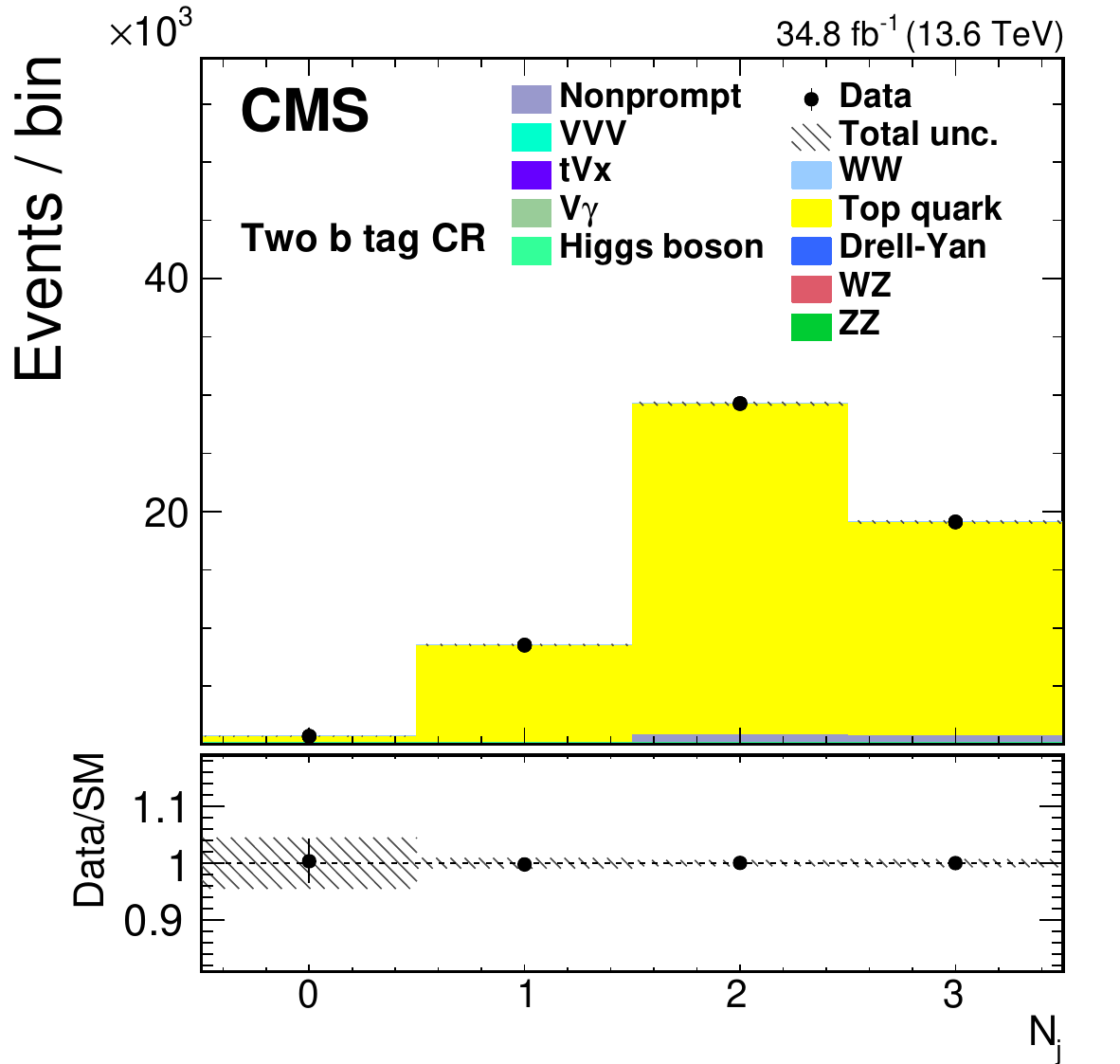}
\includegraphics[width=0.41\textwidth]{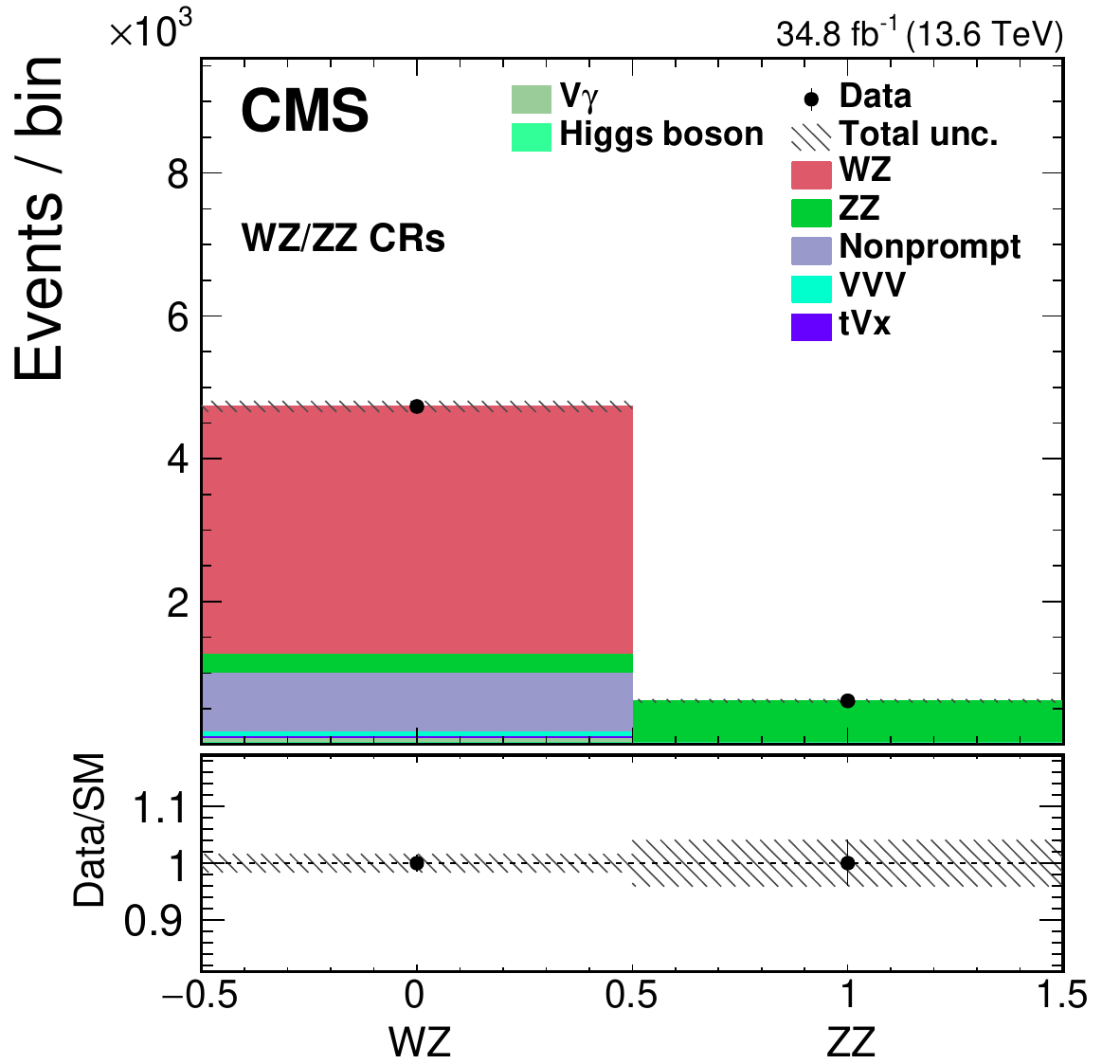}
\caption{Distributions of the number of reconstructed jets after the fit to the data in the  $\PW\PW$ SR (upper left), same-sign CR (upper right), $\PZ\to\PGt\PGt$ CR (center left), one \PQb tag CR (center right), and two \PQb tags CR (lower left), and the number of events in the $\WZ\to3\Pell\PGn$ and $\PZ\PZ\to4\Pell$ CRs (lower right). The histograms for $\tVx$ backgrounds include the contributions from $\ttbar\PV(\PV)$ processes. The histograms for $\PV\PGg$ backgrounds include the contributions from $\PW\PGg$ and $\PZ\PGg$ processes. The overflow is included in the last bin. The bottom panel in each figure shows the ratio of the number of events observed in data to that of the total SM prediction. The grey bands represent the uncertainties in the predicted yields. The vertical bars represent the statistical uncertainties in the data.}
\label{fig:wwsel_fit}
\end{figure*}

\begin{figure}[htbp!]
  \centering
  \includegraphics[width=0.49\textwidth]{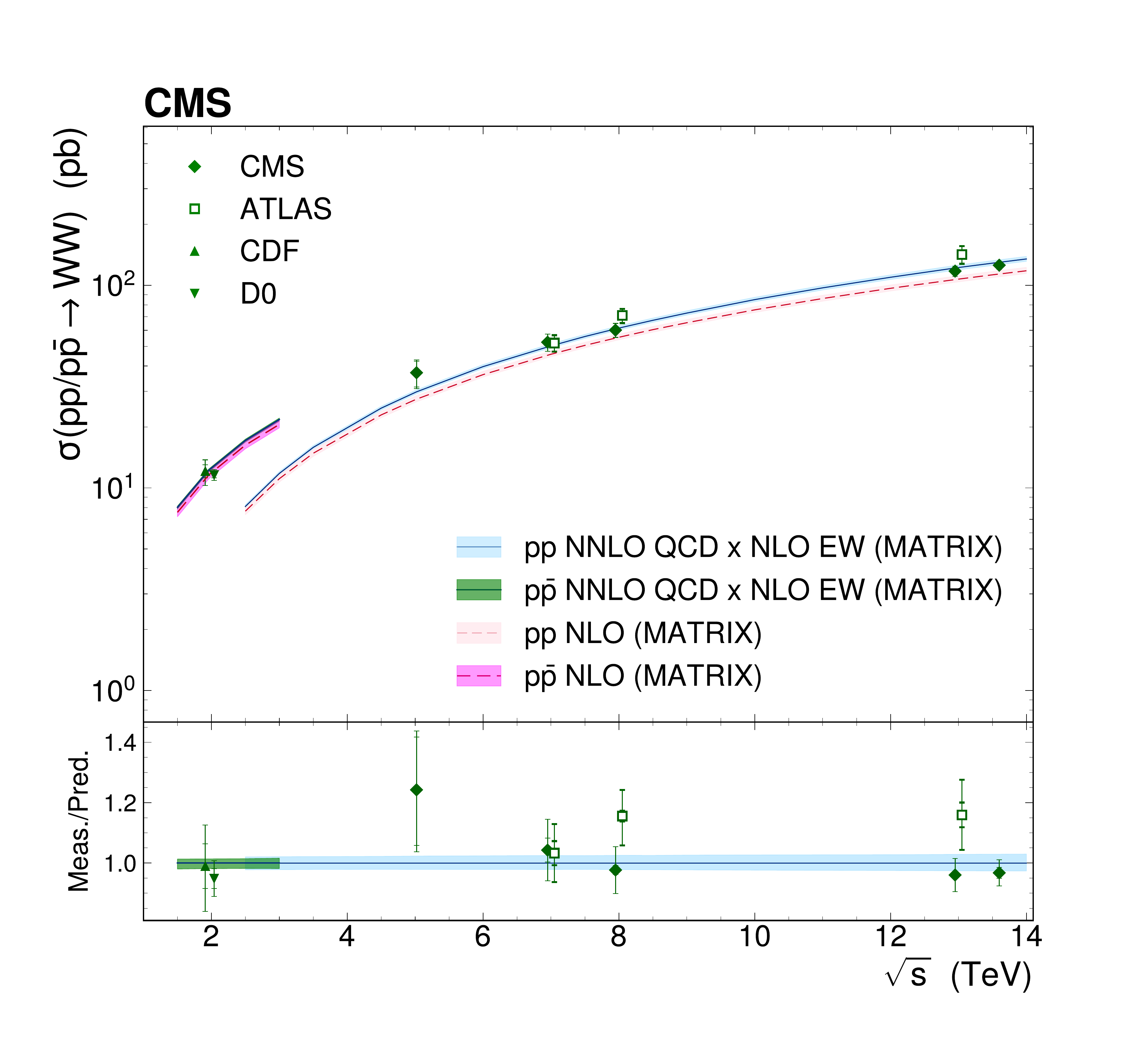}
  \caption{Results obtained in this analysis and other \WW production cross section measurements at different center-of-mass energies for the CMS~\cite{Chatrchyan:2013yaa, Khachatryan:2015sga, CMS:2020mxy}, ATLAS~\cite{ATLAS:2012mec, Aad:2016wpd, Aaboud:2017qkn}, CDF~\cite{Aaltonen:2009aa}, and D0~\cite{D0:2013bwr} Collaborations are presented, and compared with the NNLO QCD $\times$ NLO EW and NLO predictions from \MATRIX. The vertical error bars represent the uncertainty in the measured cross section.}
  \label{fig:sigmaSumm}
\end{figure}

\begin{figure*}[htbp!]
\centering
\includegraphics[width=0.80\textwidth]{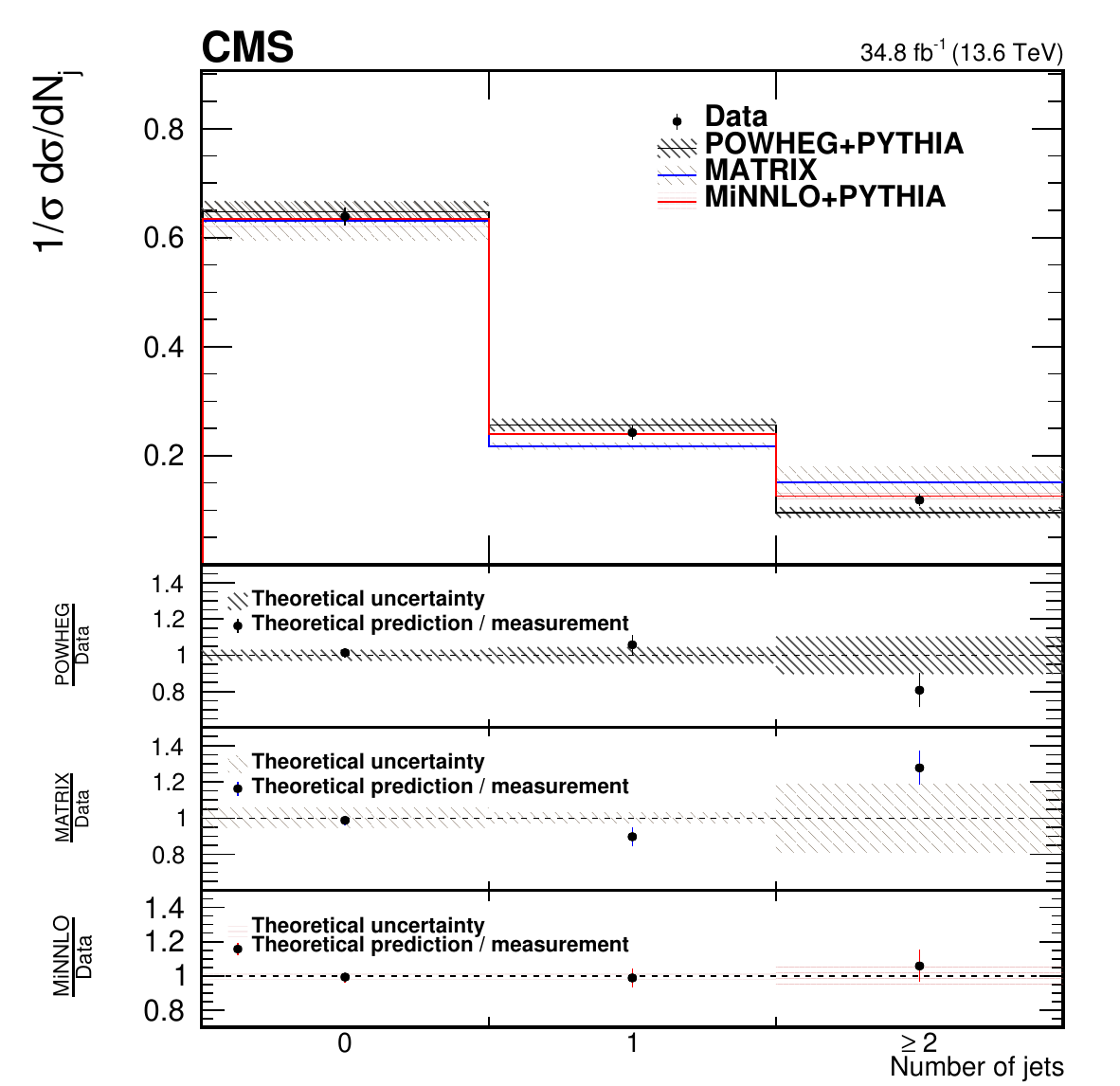}
\caption{\label{fig:njets}The upper panel shows the normalized cross section measurement for events with $\NJ=0$, 1, ${\ge}2$ jets. The filled circles represent the unfolded data results. The black, blue, and red lines represent the predictions from {\POWHEG}+\PYTHIA, \MATRIX, and \pminnlo+\PYTHIA, respectively. The lower panels show the ratio of the theoretical predictions to the measurement. In all panels, the error bars on the data points represent the total uncertainty in the measurement, and the shaded band depicts the uncertainty in the prediction.}
\end{figure*}

The data, signal, and background yields in all regions used in the analysis are shown in Tables~\ref{tab:yields_all} and~\ref{tab:yields_vv}. In the background categories, the symbol $\PV$ represents either a $\PW$ or $\PZ$ boson, and $\tVx$ represents final states with a top quark, a gauge boson, and possibly additional jets. The distributions of the number of reconstructed jets, and the number of events in the $\WZ\to3\ell\nu$ and $\PZ\PZ\to4\ell$ CRs after the fit to the data are shown in Fig.~\ref{fig:wwsel_fit}. Because of the anticorrelation between the yields of the signal and background processes, the total uncertainty in the different regions is smaller than the uncertainties in some of the individual contributions. The goodness-of-fit using the saturated test statistic~\cite{CMS:2024onh,Cousins:2018tiz} gives a p-value of 26\% that characterizes the compatibility of the observed data and the statistical model. After the fit to data, all nuisance parameters differ by less than one standard deviation with respect to their initial predicted values. The observed normalization of the most important background, top quark production, is consistent with dedicated cross section measurements~\cite{ATLAS:2023slx,CMS:2023qyl}. The measured inclusive cross section based on the simultaneous fit of all event categories is $125.7 \pm 2.3\stat \pm 4.8\syst \pm 1.8\lum\pb = 125.7\pm 5.6\pb$, which is consistent with the theoretical prediction within uncertainties. A summary of the inclusive fiducial cross section and normalized cross sections obtained in the analysis is shown in Table~\ref{tab:signalstrengths}. Figure~\ref{fig:sigmaSumm} presents a summary of  \WW production cross section measurements at different center-of-mass energies and a comparison with fixed-order predictions produced via the \MATRIX\ framework~\cite{Grazzini:2017mhc,Kallweit:2018pee}. Results are consistent with the theoretical predictions across all center-of-mass energies. The normalized fiducial cross section measurement as a function of $\NJ$ is shown in Fig.~\ref{fig:njets}. The measurement is compared to predictions with \POWHEG{+}\PYTHIA, NNLO QCD $\times$ NLO EW predictions using \MATRIX\ calculations, and NNLO in QCD predictions using \pminnlo~\cite{Monni:2019whf,Monni:2020nks}~v2 interfaced with \PYTHIA. The measured fiducial cross sections agree with the theoretical predictions within uncertainties. Nevertheless, the predictions using \pminnlo seem to give a sizable improvement with respect to older predictions. Tabulated results are provided in HEPData~\cite{HEPData}. The statistical model for this analysis is provided in~\cite{StatModel}.

\begin{table*}[htbp!]
\centering
\topcaption{Data, signal, and background yields combining all reconstructed jet multiplicity bins. The combination of the statistical and systematic uncertainties is shown. The expected yields are shown with their best fit normalizations from the simultaneous fit to data.\label{tab:yields_all}}
\newcolumntype{x}{D{,}{\,\pm\,}{5.3}}
\begin{tabular}{lxxxxx{c}@{\hspace*{5pt}}x}
\hline
                     & \multicolumn{1}{c}{$\PW\PW$ SR}         & \multicolumn{1}{c}{Same-sign CR}         & \multicolumn{1}{c}{$\PZ\to\tau\tau$ CR}   & \multicolumn{1}{c}{One \PQb tag CR}     & \multicolumn{1}{c}{Two \PQb tags CR}  \\
\hline
                Data & \multicolumn{1}{c}{43\,898} & \multicolumn{1}{c}{3\,456} & \multicolumn{1}{c}{56\,551} & \multicolumn{1}{c}{68\,656} & \multicolumn{1}{c}{57\,617} \\[\cmsTabSkip]
            $\PW\PW$ & 16\,030 ,   160 &     80.9 ,   5.0 &   2\,648 ,    37   &  2\,128 ,    27 &     227 ,  11 \\
           Top quark & 19\,900 ,   130 &     87.9 ,   3.7 & 1\,131.6 ,   8.8   & 63\,450 ,   210 & 55\,550 , 210 \\
    $\PZ\to\tau\tau$ &  2\,124 ,    31 &     57.1 ,   4.1 &  45\,610 ,   320   &   226.0 ,   8.6 &    19.5 , 2.3 \\
               $\WZ$ &   487.7 ,   8.6 &    512.1 ,   9.6 & 	97.6 ,   2.1   &    97.1 ,   2.1 &    11.8 , 0.6 \\
               $\ZZ$ &    37.1 ,   0.8 &     33.6 ,   0.7 & 	66.0 ,   1.6   &     6.9 ,   0.2 &     1.0 , 0.0 \\
           Nonprompt &  4\,880 ,   140 &   2\,390 ,    53 &   6\,580 ,   280   &  2\,670 ,   100 &  1\,695 ,  98 \\
         $\PV\PV\PV$ &    76.0 ,   1.6 &     25.8 ,   0.6 & 	 4.7 ,   0.1   &    33.7 ,   1.1 &     8.6 , 0.4 \\
              $\tVx$ &    10.7 ,   0.5 &      8.7 ,   1.1 & 	 0.7 ,   0.1   &    44.1 ,   1.4 &    52.0 , 1.6 \\
         $\PV\gamma$ &   226.5 ,   7.7 &    233.3 ,   7.5 & 	68.8 ,   3.4   &    43.4 ,   3.4 &     3.0 , 0.4 \\
         Higgs boson &    90.2 ,   6.4 &     27.2 ,   1.7 & 	 345 ,    32   &    29.4 ,   1.7 &    20.9 , 1.2 \\[\cmsTabSkip]
               Total & 43\,860 ,   180 &   3\,457 ,    53 &  56\,550 ,   240   & 68\,720 ,   220 & 57\,590 , 230 \\
\hline
\end{tabular}
\end{table*}

\begin{table*}[htbp]
\centering
\topcaption{Data, signal and background yields in events in the \WZ and \ZZ CRs. The combination of the statistical and systematic uncertainties is shown. The expected yields are shown with their best fit normalizations from the simultaneous fit to data.}
\label{tab:yields_vv}
\newcolumntype{x}{D{,}{\,\pm\,}{4.3}}
\begin{tabular}{lxx{c}@{\hspace*{5pt}}x}
\hline
                       & \multicolumn{1}{c}{$\WZ$ CR} & \multicolumn{1}{c}{$\ZZ$ CR} \\
\hline
                Data &    \multicolumn{1}{c}{4\,732} & \multicolumn{1}{c}{610}         \\[\cmsTabSkip]
               $\WZ$ &  3\,470 ,    85  	     &    0.9 ,   0.1		       \\
               $\ZZ$ &     271 ,    20  	     &    599 ,    25		       \\
           Nonprompt &     819 ,    80  	     &     \multicolumn{1}{c}{$<1$}    \\
         $\PV\PV\PV$ &    60.5 ,   2.9  	     &    5.4 ,   0.2		       \\
              $\tVx$ &    26.0 ,   2.3  	     &    2.3 ,   0.2		       \\
         $\PV\gamma$ &    55.8 ,   6.0  	     &     \multicolumn{1}{c}{$<1$}    \\
         Higgs boson &    28.4 ,   1.9  	     &    2.4 ,   0.6		       \\[\cmsTabSkip]
               Total &  4\,729 ,    76  	     &    610 ,    25 \\
\hline
\end{tabular}
\end{table*}

\begin{table*}[htbp!]
\centering
\topcaption{\label{tab:signalstrengths}
Inclusive fiducial cross section and normalized cross sections for events with $\NJ=0, 1, {\ge}2$ jets. The uncertainty listed is the total uncertainty obtained from the fit to the yields. The expected predictions are obtained from {\POWHEG}+\PYTHIA. In parentheses, the split of systematic and statistical uncertainties are reported.}
\begin{tabular}{lcc}
\hline
Observable & Expected & Observed \\
\hline
Cross section (fb)   & $  812 \pm    34\,(   31,   15)$ & $  813 \pm    35\,(   32,   15)$ \\
      0-jet fraction & $0.648 \pm 0.015\,(0.012, 0.009)$ & $0.640 \pm 0.016\,(0.013, 0.009)$ \\
      1-jet fraction & $0.256 \pm 0.013\,(0.008, 0.010)$ & $0.243 \pm 0.013\,(0.009, 0.010)$ \\
$\geq$2-jet fraction & $0.096 \pm 0.011\,(0.008, 0.008)$ & $0.119 \pm 0.011\,(0.008, 0.008)$ \\
\hline
\end{tabular}
\end{table*}

\section{Summary}

First measurements at $\sqrt{s}=13.6\TeV$ of the \WW production cross section in proton-proton collisions have been presented. The data used in this study were collected with the CMS detector at the CERN LHC in 2022, and correspond to an integrated luminosity of 34.8\fbinv. Events were selected by requiring one electron and one muon of opposite charge. A maximum likelihood fit has been performed in event signal- and background-enriched categories defined by the flavour and charge of the leptons, the number of jets, and number of jets identified as originating from \PQb quarks. An inclusive  \WW production cross section of $125.7 \pm 2.3\stat \pm 4.8\syst \pm 1.8\lum\pb = 125.7\pm 5.6\pb$ is measured, in agreement with the standard model prediction. The overall sensitivity is about 25\% better than previous CMS measurements at $\sqrt{s}=13\TeV$ with a similar integrated luminosity, because of several reduced experimental uncertainties and the improved fit strategy. Cross sections have also been reported in a fiducial region close to that of the detector acceptance, both inclusively and differentially, as a function of the jet multiplicity in the event. For the first time in proton-proton collisions, \WW events with zero, one, and at least two jets are studied simultaneously and compared with recent theoretical predictions.

\begin{acknowledgments}
We congratulate our colleagues in the CERN accelerator departments for the excellent performance of the LHC and thank the technical and administrative staffs at CERN and at other CMS institutes for their contributions to the success of the CMS effort. In addition, we gratefully acknowledge the computing centres and personnel of the Worldwide LHC Computing Grid and other centres for delivering so effectively the computing infrastructure essential to our analyses. Finally, we acknowledge the enduring support for the construction and operation of the LHC, the CMS detector, and the supporting computing infrastructure provided by the following funding agencies: SC (Armenia), BMBWF and FWF (Austria); FNRS and FWO (Belgium); CNPq, CAPES, FAPERJ, FAPERGS, and FAPESP (Brazil); MES and BNSF (Bulgaria); CERN; CAS, MoST, and NSFC (China); MINCIENCIAS (Colombia); MSES and CSF (Croatia); RIF (Cyprus); SENESCYT (Ecuador); ERC PRG, RVTT3 and MoER TK202 (Estonia); Academy of Finland, MEC, and HIP (Finland); CEA and CNRS/IN2P3 (France); SRNSF (Georgia); BMBF, DFG, and HGF (Germany); GSRI (Greece); NKFIH (Hungary); DAE and DST (India); IPM (Iran); SFI (Ireland); INFN (Italy); MSIP and NRF (Republic of Korea); MES (Latvia); LMTLT (Lithuania); MOE and UM (Malaysia); BUAP, CINVESTAV, CONACYT, LNS, SEP, and UASLP-FAI (Mexico); MOS (Montenegro); MBIE (New Zealand); PAEC (Pakistan); MES and NSC (Poland); FCT (Portugal);  MESTD (Serbia); MCIN/AEI and PCTI (Spain); MOSTR (Sri Lanka); Swiss Funding Agencies (Switzerland); MST (Taipei); MHESI and NSTDA (Thailand); TUBITAK and TENMAK (Turkey); NASU (Ukraine); STFC (United Kingdom); DOE and NSF (USA).

\hyphenation{Rachada-pisek} Individuals have received support from the Marie-Curie programme and the European Research Council and Horizon 2020 Grant, contract Nos.\ 675440, 724704, 752730, 758316, 765710, 824093, 101115353, 101002207, and COST Action CA16108 (European Union); the Leventis Foundation; the Alfred P.\ Sloan Foundation; the Alexander von Humboldt Foundation; the Science Committee, project no. 22rl-037 (Armenia); the Belgian Federal Science Policy Office; the Fonds pour la Formation \`a la Recherche dans l'Industrie et dans l'Agriculture (FRIA-Belgium); the Agentschap voor Innovatie door Wetenschap en Technologie (IWT-Belgium); the F.R.S.-FNRS and FWO (Belgium) under the ``Excellence of Science -- EOS" -- be.h project n.\ 30820817; the Beijing Municipal Science \& Technology Commission, No. Z191100007219010 and Fundamental Research Funds for the Central Universities (China); the Ministry of Education, Youth and Sports (MEYS) of the Czech Republic; the Shota Rustaveli National Science Foundation, grant FR-22-985 (Georgia); the Deutsche Forschungsgemeinschaft (DFG), under Germany's Excellence Strategy -- EXC 2121 ``Quantum Universe" -- 390833306, and under project number 400140256 - GRK2497; the Hellenic Foundation for Research and Innovation (HFRI), Project Number 2288 (Greece); the Hungarian Academy of Sciences, the New National Excellence Program - \'UNKP, the NKFIH research grants K 131991, K 133046, K 138136, K 143460, K 143477, K 146913, K 146914, K 147048, 2020-2.2.1-ED-2021-00181, and TKP2021-NKTA-64 (Hungary); the Council of Science and Industrial Research, India; ICSC -- National Research Centre for High Performance Computing, Big Data and Quantum Computing and FAIR -- Future Artificial Intelligence Research, funded by the NextGenerationEU program (Italy); the Latvian Council of Science; the Ministry of Education and Science, project no. 2022/WK/14, and the National Science Center, contracts Opus 2021/41/B/ST2/01369 and 2021/43/B/ST2/01552 (Poland); the Funda\c{c}\~ao para a Ci\^encia e a Tecnologia, grant CEECIND/01334/2018 (Portugal); the National Priorities Research Program by Qatar National Research Fund; MCIN/AEI/10.13039/501100011033, ERDF ``a way of making Europe", and the Programa Estatal de Fomento de la Investigaci{\'o}n Cient{\'i}fica y T{\'e}cnica de Excelencia Mar\'{\i}a de Maeztu, grant MDM-2017-0765 and Programa Severo Ochoa del Principado de Asturias (Spain); the Chulalongkorn Academic into Its 2nd Century Project Advancement Project, and the National Science, Research and Innovation Fund via the Program Management Unit for Human Resources \& Institutional Development, Research and Innovation, grant B37G660013 (Thailand); the Kavli Foundation; the Nvidia Corporation; the SuperMicro Corporation; the Welch Foundation, contract C-1845; and the Weston Havens Foundation (USA).
\end{acknowledgments}

\bibliography{auto_generated}
\cleardoublepage \appendix\section{The CMS Collaboration \label{app:collab}}\begin{sloppypar}\hyphenpenalty=5000\widowpenalty=500\clubpenalty=5000
\cmsinstitute{Yerevan Physics Institute, Yerevan, Armenia}
{\tolerance=6000
A.~Hayrapetyan, A.~Tumasyan\cmsAuthorMark{1}\cmsorcid{0009-0000-0684-6742}
\par}
\cmsinstitute{Institut f\"{u}r Hochenergiephysik, Vienna, Austria}
{\tolerance=6000
W.~Adam\cmsorcid{0000-0001-9099-4341}, J.W.~Andrejkovic, T.~Bergauer\cmsorcid{0000-0002-5786-0293}, S.~Chatterjee\cmsorcid{0000-0003-2660-0349}, K.~Damanakis\cmsorcid{0000-0001-5389-2872}, M.~Dragicevic\cmsorcid{0000-0003-1967-6783}, P.S.~Hussain\cmsorcid{0000-0002-4825-5278}, M.~Jeitler\cmsAuthorMark{2}\cmsorcid{0000-0002-5141-9560}, N.~Krammer\cmsorcid{0000-0002-0548-0985}, A.~Li\cmsorcid{0000-0002-4547-116X}, D.~Liko\cmsorcid{0000-0002-3380-473X}, I.~Mikulec\cmsorcid{0000-0003-0385-2746}, J.~Schieck\cmsAuthorMark{2}\cmsorcid{0000-0002-1058-8093}, R.~Sch\"{o}fbeck\cmsorcid{0000-0002-2332-8784}, D.~Schwarz\cmsorcid{0000-0002-3821-7331}, M.~Sonawane\cmsorcid{0000-0003-0510-7010}, W.~Waltenberger\cmsorcid{0000-0002-6215-7228}, C.-E.~Wulz\cmsAuthorMark{2}\cmsorcid{0000-0001-9226-5812}
% ...
\par}
\vskip\cmsinstskip
\dag:~Deceased\\
$^{1}$Also at Yerevan State University, Yerevan, Armenia\\
$^{2}$Also at TU Wien, Vienna, Austria\\
% ...
\end{sloppypar}
\end{document}